\documentclass[aps,prb,twocolumn,groupedaddress]{revtex4}
\usepackage{graphicx}
\begin{document}


\title{Pressure-induced unconventional superconductivity near a quantum critical point in CaFe$_2$As$_2$}

\author{S. Kawasaki$^1$}
\author{T. Tabuchi$^1$}
\author{X. F. Wang$^2$}
\author{X. H. Chen$^2$}
\author{Guo-qing Zheng$^{1,3}$}

\affiliation{$^1$Department of Physics, Okayama University, Okayama 700-8530, Japan} 
\affiliation{$^2$Hefei National Laboratory for Physical Sciences at Microscale and Department of Physics, University of Science and Technology of China, Hefei, Anhui 230026, China}
\affiliation{$^3$Institute of Physics and Beijing National Laboratory
for Condensed Matter Physics, Chinese Academy of Sciences, Beijing 100190, China}





\begin{abstract}
$^{75}$As-zero-field nuclear magnetic resonance (NMR) and nuclear quadrupole resonance (NQR) measurements are performed on CaFe$_2$As$_2$ under pressure. At $P$ = 4.7 and 10.8 kbar, the temperature dependences of nuclear-spin-lattice relaxation rate (1/$T_1$) measured in the tetragonal phase show no coherence peak just below $T_c(P)$ and decrease with decreasing temperature. The superconductivity is gapless at $P$ = 4.7 kbar but evolves to that with multiple gaps at $P$ = 10.8 kbar.  We find that the superconductivity appears near a quantum critical point under pressures in the range 4.7 kbar $\le P \le$ 10.8 kbar. Both electron correlation and superconductivity disappear in the collapsed tetragonal phase. A systematic study under pressure indicates that electron correlations play a vital role in forming Cooper pairs in this compound. 

\end{abstract}


\maketitle


\section{Introduction}
Since the discovery of high-transition temperature ($T_c$) superconductivity in LaFeAsO$_{1-x}$F$_x$ at $T_c$ = 26 K\cite{Kamihara}, iron pnictides have become one of the most fascinating research areas in condensed-matter physics. The electron-doping suppresses structural and magnetic phase transitions in undoped ReFeAsO and superconductivity appears near the border of magnetism \cite{Kamihara} as seen in the high-$T_c$ cuprates. After the discovery of superconductivity in ReFeAsO, the high-$T_c$ superconductivity has also been found in a ThCr$_2$Si$_2$-type structure, BaFe$_2$As$_2$ by replacing Ba by K as hole doping\cite{Rotter}. One of the most remarkable features in these iron-pnictide superconductors is their superconducting gap structure. Previous nuclear-magnetic resonance (NMR) and nuclear-quadrupole resonance (NQR) measurements on ReFeAsOF and Ba$_{0.7}$K$_{0.3}$Fe$_2$As$_2$ consistently found multiple gap superconductivity\cite{MatanoPr,KawasakiLa,MatanoBa122}. These observations were also confirmed by angle-resolved photoemission spectroscopy. \cite{Ding} The multiple gap feature is believed to be relevant to their multiple electronic band structure.\cite{Singh} 
On the other hand, the mechanism of the Cooper pair formation in iron pnictide is still unclear. Since the superconductivity in iron-pnictides is induced by chemical doping, systematic investigation of the relationship between electron correlations and superconductivity has been difficult.

Recent discoveries of pressure-induced superconductivity in RFe$_2$As$_2$ (R = Ca, Sr, and Ba) provide a new route to investigate superconductivity in iron pnictide\cite{Lonzarich,Canfield,Thompson,Kotegawa}. In RFe$_2$As$_2$, the parent compounds also show structural transition from a tetragonal (tetra.) to orthorhombic (orth.) structure with antiferromagnetic order\cite{KitagawaBa,KitagawaSr,CurroCa}. In BaFe$_2$As$_2$ and SrFe$_2$As$_2$, the structural phase transition and antiferromagnetic orders are both suppressed by  pressure, and superconductivity was found around the critical pressure, $P_c$ = 40 - 60 kbar, with $T_c$ $\sim$ 30 K.\cite{Lonzarich,Kotegawa} On the other hand, pressure-induced superconductivity in CaFe$_2$As$_2$ has been observed with lower $P_c$ $\sim$ 5 kbar and lower $T_c(P)$ $\sim$ 10K.\cite{Canfield,Thompson}  The most strikingly different feature in CaFe$_2$As$_2$ is the occurrence of another structural transition under pressure. Above $P$ $\sim$ 5 kbar, normal state tetra. phase changes to a collapsed tetragonal (c-tetra.) phase with drastic reduction in both the unit cell volume (5\%) and the $c/a$ ratio (11\%).\cite{Kreyssig,Goldman}  Notably, when tetra. phase collapses, superconductivity disappears.\cite{Yu}  Since these structural phase transitions are sensitive to external pressure, the detailed information about pressure-induced superconductivity in CaFe$_2$As$_2$ is still unknown.

In this paper, we report results of zero-field (ZF) NMR and NQR study in CaFe$_2$As$_2$.  At $P$ = 4.7 and 10.8 kbar, pressure-induced superconductivity in the tetra. phase is confirmed by ac-susceptibility and nuclear-spin lattice relaxation time ($T_1$) measurements. The temperature dependences of 1/$T_1$ show no coherence peak just below $T_c(P)$. Below $T_c(P)$, the temperature dependences of $1/T_1$ indicate the unconventional nature of pressure-induced superconductivity in CaFe$_2$As$_2$. The systematic measurements indicate electron correlations play a vital role in inducing unconventional superconductivity in this compound.

\section{Experimental procedures}
The single crystals of CaFe$_2$As$_2$ are grown by a self-flux method and crushed into coarse powder for $^{75}$As ($I$ = 3/2, $\gamma$ = 7.292 MHz/T) ZF-NMR/NQR measurements under pressure.
The pressure was applied by utilizing a NiCrAl/BeCu piston-cylinder type cell filled with Daphne 7373 as the pressure-transmitting medium\cite{Murata}. The pressure at low temperatures was determined from the pressure dependence of the $T_{c}$ values of Sn metal measured by a conventional four-terminal method. The temperature dependence of ac-susceptibility is measured using an $in$-$situ$ NMR/NQR coil.  The ZF-NMR/NQR spectra were taken by changing rf frequency and recording the spin echo intensity step by step.  The value of $T_1$ was extracted by fitting the nuclear magnetization obtained by recording the spin echo intensity after the saturation pulse. 

\begin{figure}[h]
\begin{center}
\includegraphics[width=6cm]{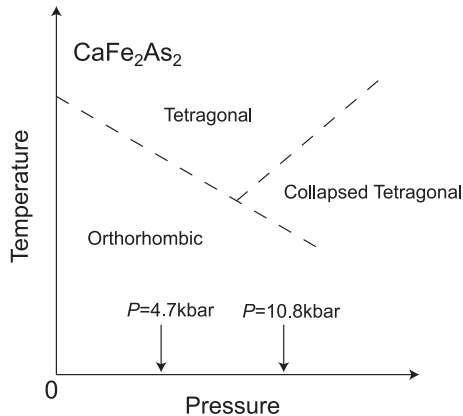}
\caption{\label{fig:t1t} Schematic phase diagram for structural phase transitions of CaFe$_2$As$_2$ under pressure (see text). }
\end{center}
\end{figure}

Figure 1 shows the schematic phase diagram of CaFe$_2$As$_2$ under pressure taken from the literature\cite{Kreyssig,Goldman,Yu,Baek}. Dashed lines indicate the first-order structural phase transitions, respectively. Arrows indicate the pressure at which the present experiments have been performed.


\section{Pressure dependence of zero-field NMR and NQR spectra}
\begin{figure}[!]
\begin{center}
\includegraphics[width=8.5cm]{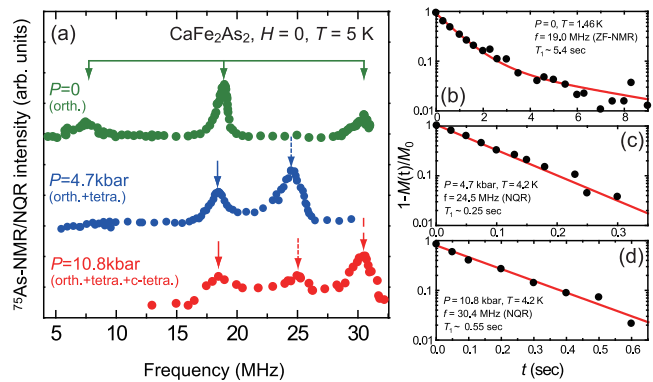}

\caption{\label{fig:t1t} (Color online) (a) Pressure dependence of $^{75}$As-NMR/NQR spectra for CaFe$_2$As$_2$ measured at $T$ = 5 K and $H$ = 0. Solid arrows indicate the $^{75}$As ZF-NMR spectrum which comes from the orth. phase below $T_N$. Dotted and dashed arrows indicate NQR spectra at the tetra. and c-tetra. phases, respectively. Typical data sets of nuclear recovery curves measured at (b) orth. phase (ZF-NMR), (c) tetra. phase (NQR), and (d) c-tetra. phase (NQR), respectively. Solid curves are theoretical fittings to obtain $T_1$. (see text) }
\end{center}
\end{figure}

Figures 2 (a) shows the pressure dependence of ZF-NMR and NQR spectrum measured at $T$ = 5 K and $P$ = 0, 4.7, and 10.8 kbar, respectively. At $P$ = 0, three $^{75}$As-NMR lines are observed due to the internal magnetic field ($H_{int}$) induced by the Fe ordered moment below $T_N$, which comes from the orth. phase. Actually, as seen in Figs. 2 (b), the nuclear magnetization recovery curve measured at 19 MHz is well fitted by the theoretical curve for NMR ($-1/2\leftrightarrow +1/2$ transition) which is given by 1-$M(t)$/$M_0$ = 0.1$\exp$(-$t/T_1$)+0.9$\exp$(-6$t/T_1$), where $M_0$ and $M(t)$ are the nuclear magnetization at the thermal equilibrium and at a time $t$ after saturating pulse, respectively.  Assuming that both $H_{int}$ at the As site and $\nu_Q$ for the As nuclei are along the $c$-axis direction, $H_{int}$ = 2.6 T and $\nu_Q$ = 12 MHz are obtained. Here, nuclear spin Hamiltonian is given as $\mathcal{H}_{AFM}=-\gamma\hbar\vec{I}\cdot\vec{H}_{int} + (h \nu_{Q}/6)[3{I_z}^2-I(I+1)]$. These parameters are in good agreement with previous As-NMR experiments on single crystalline CaFe$_2$As$_2$\cite{CurroCa}.

 While the ground state of CaFe$_2$As$_2$ at ambient pressure is in a single orth. phase, a phase separation is observed under pressure due to the first-order transition\cite{CurroCa,Goldman} and pressure distribution\cite{Yu}. As seen in Figs.2(a), at $P$ = 4.7 kbar, a phase separation between orth. and tetra. phases is observed as the ground state. The peak around 18 MHz is due to the central transition ($-1/2\leftrightarrow +1/2$ transition) for ZF-NMR of the orth. phase as observed at $P$ = 0. However, the satellite peaks which are clearly observed at ambient pressure, due to nuclear quadrupole interaction is not observed, indicating an increase of $\nu_Q$ for the orth. phase under pressure. On the other hand, another peak appears around 24 MHz. Since the nuclear magnetization recovery curve measured at 24 MHz is well fitted by the nuclear magnetization recovery curve for $^{75}$As-NQR ($\pm 1/2\leftrightarrow \pm 3/2$ transition) given by the single exponential 1-$M(t)$/$M_0$ = $\exp$(-3$t/T_1$), as seen in Figs. 2 (c), we assigned this peak as coming from the tetra. phase which survives due to a pressure distribution. We have also confirmed this assignment by measuring As-NMR spectrum at $P$ = 5.0 kbar (not shown). Notably, it has been reported that the structural transition from tetra. to orth. under pressure is accompanied by a phase separation in a certain temperature range until a single otrh. phase is established as the magnetic ground state\cite{Goldman,Baek}. Since the present experiment is performed using coarse powdered single crystals, the local pressure distribution may cause the tetra. phase to coexist  with the orth. phase even at the ground state.      

As pressure reaches $P$ = 10.8 kbar another structural transition from tetra. to c-tetra. occurs\cite{Kreyssig,Goldman}, the NMR signal around 18 MHz and the NQR signal from tetra. around 25 MHz are still observed. In addition, a new peak appears around 30.4 MHz. As seen in Figs.2(d), since the nuclear magnetization recovery curve at 30.4 MHz indicates this new peak is also from NQR, we assigned that this peak as coming from the c-tetra. phase. Due to a pressure distribution, phase separation among orth., tetra., and c-tetra. is realized at $P$ = 10.8 kbar. Notably, an As-NQR frequency $\nu_Q$ probes the electric-field gradient (EFG) generated by the charge distribution surrounding the As site. The larger $\nu_Q$ for c-tetra. than tetra. phase is reasonable since the unit cell volume collapses in c-tetra. However, these pressure and structural dependences of $\nu_Q$ do not scale with the known unit cell volume for each phases\cite{Kreyssig}. This may be because the local charge distribution around the As site  also contributes to EFG in addition to the lattice contribution.  Since the NMR/NQR peaks are clearly separated even at $P$ = 10.8 kbar and the recovery curves measured at each phases are of a single $T_1$ component, we suggest that these phase separations under pressure are induced by a local pressure distribution and  not by the sample inhomogeneity. Such a pressure distribution allowed us to investigate the pressure dependence of the electronic properties in each phase of CaFe$_2$As$_2$.

From the NMR/NQR spectra, the volume fraction of orth.:tetra.=54\%:46\% and orth.:tetra.:c-tetra.=45\%:18\%:37\% are estimated for $P$ = 4.7 and 10.8 kbar, respectively. It is clear that the effect of pressure distribution on the evolution of the ground states in CaFe$_2$As$_2$ is larger than a previous NMR study under pressure using a large single crystal\cite{Baek}.   

\section{Pressure-induced superconductivity in CaFe$_2$As$_2$}
Figure 3 shows temperature dependence of ac-susceptibility measured using the $in$-$situ$ NMR/NQR coil. The pressure-induced superconducting transitions at $T_c(P)$ = 3.9 and 4.1 K at $P$ = 4.7 and 10.8 kbar are clearly observed. Although the $T_c(P)$s are relatively lower, the superconducting  transitions are much sharper than previous reports\cite{Baek,Hanoh}. 
 
\begin{figure}[h]
\begin{center}
\includegraphics[width=7cm]{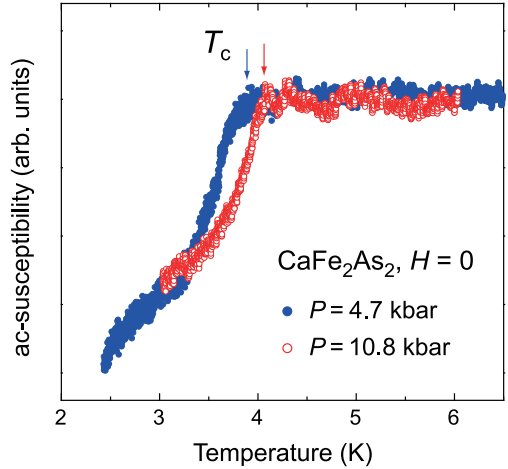}

\caption{\label{fig:t1t} (Color online) Temperature dependence of ac-susceptibility measured using $in$-$situ$ NMR/NQR coil at $P$ = 4.7 and 10.8 kbar. Arrows indicate $T_c(P)$. }
\end{center}
\end{figure}

\section{Evolution of electronic property in CaFe$_2$As$_2$ under pressure}
To investigate the evolution of the electronic  properties in CaFe$_2$As$_2$ under pressure, we measured the $^{75}$As nuclear spin lattice relaxation time ($T_1$) at each phase. Figs. 4 shows the temperature dependence of $1/T_1$ divided by temperature ($1/T_1T$) well below the structural transitions. In the orth. phases all of the  data show a $1/T_1T$ = constant behavior, which is characteristic of a Fermi-liquid state. These results are consistent with previous NMR results on (Ba, Sr)Fe$_2$As$_2$\cite{KitagawaBa,KitagawaSr}, indicating that a small Fermi surface remains below antiferromagnetic order. Due to the large $H_{int}$ $\sim$ 2.5T in the orth. phase, the coexistence of antiferromagnetism and superconductivity, which has frequently been observed in heavy-fermion compounds\cite{Kitaoka}, could not be confirmed. 
As discussed later, pressure-induced superconductivity is clearly observed as a reduction of $1/T_1T$  for the tetra. phases at which the onset of diamagnetism is observed at $P$ = 4.7 and 10.8 kbar, respectively.

In the normal state in the tetra. phases, $1/T_1T$ increases with decreasing temperature, which indicates that the antiferromagnetic correlation develops down to $T_c(P)$.  To analyze the temperature dependence of $1/T_1T$ above $T_c(P)$, we employed the model for a weakly antiferromagnetically correlated metal, 1/$T_1T$ = const. + $C/(T+\theta)$ \cite{Moriya}.  Here, the first term describes the contribution from the density of states (DOS) at the Fermi level, and the second term describes the contribution from the antiferromagnetic wave vector $Q$. As shown by the  solid curves in Fig. 4, the temperature dependences of 1/$T_1T$ for tetra. phases are well fitted by this model; 1/$T_1T$ = 0.48 + $4.4/(T+\theta)$ with $\theta$ = 5.4$\pm$2.3 K  for $P$ = 4.7 kbar and 1/$T_1T$ = 0.57 + $5.2/(T+\theta)$ with $\theta$ = 6.0$\pm$1.1 K for $P$ = 10.8 kbar, respectively. Surprisingly, the values of $\theta$, which is a measure of the distance to an antiferromagnetic quantum critical point (QCP), are not only one order of magnitude smaller than $\theta$ = 39 K observed in LaFeAsO$_{0.92}$F$_{0.08}$ ($T_c$ = 23 K)\cite{KawasakiLa}, but also comparable to that observed in unconventional superconductors in strongly correlated electron systems\cite{ZhengIr,KawasakiCeRhIrIn5,Kusano}. This indicates that superconductivity in CaFe$_2$As$_2$ is induced near an antiferromagnetic QCP. Since the value of $\theta$ is insensitive to pressure, the present results indicate that the quantum criticality in the tetra. phase is robust against pressure. This may be the reason why a robust superconducting dome was observed under pressure\cite{Canfield,Thompson}. Such a situation is somewhat different from heavy fermion superconductivity around QCP, at which both $T_c$ and electron correlations are enhanced.\cite{KawasakiCeRhIrIn5}
 On the other hand, both the DOS at the Fermi level and the value of C in the  antiferromagnetic correlation slightly increase with increasing pressure. The small increase of $T_c$, from 3.9 K at 4.7 kbar to 4.1 K at 10.8 kbar, may be due to this small increase of both the DOS at the Fermi level and the  antiferromagnetic correlations. To describe the detailed relationship between QCP and superconductivity in CaFe$_2$As$_2$, further systematic measurements under pressure are in progress.   
 
\begin{figure}[h]
\begin{center}
\includegraphics[width=7.5cm]{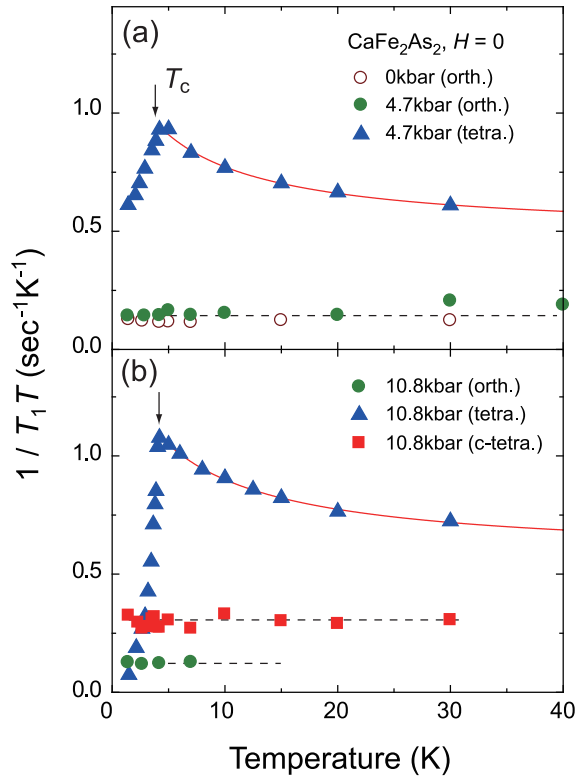}
\caption{\label{fig:t1t} (Color online) Temperature dependence of $1/T_1T$ below $T$ = 40 K in each phase at (a) $P$ = 0 and 4.7 kbar and (b) 10.8 kbar. Solid arrows indicate $T_c(P)$. Dotted lines indicates the relation of $1/T_1T$ = constant.  The solid curves indicate relation, 1/$T_1T$ = const. + $C/(T+\theta)$. (see text)   }
\end{center}
\end{figure}

The most important result is the difference of $1/T_1T$ between the tetra. and c-tetra. phases at $P$ = 10.8 kbar. As seen in Figs. 4(b), $1/T_1T$ = constant behavior is established even below $T_c$, indicating that the electron correlation and also superconductivity disappear in the c-tetra. phase. Importantly, recent electronic band structure calculations for CaFe$_2$As$_2$ have shown that the tetra. phase has the multiple band structure  seen in other iron-pnictides\cite{Ding}, whereas the multiband nature along the $\Gamma$-M direction vanishes when it collapses.\cite{LonzarichBandcalc,Yildirim} It is thus suggested that the candidate for the antiferromagnetic wave vector $Q$ observed in the tetra. phases and the driving force of the Cooper pair formation in CaFe$_2$As$_2$ is the interband correlations, which has been suggested as the origin for the spin-density-wave order in the LaFeAsOF superconductor.\cite{Ma}

\section{Novel superconductivity in CaFe$_2$As$_2$}
\begin{figure}[h]
\begin{center}
\includegraphics[width=7.5cm]{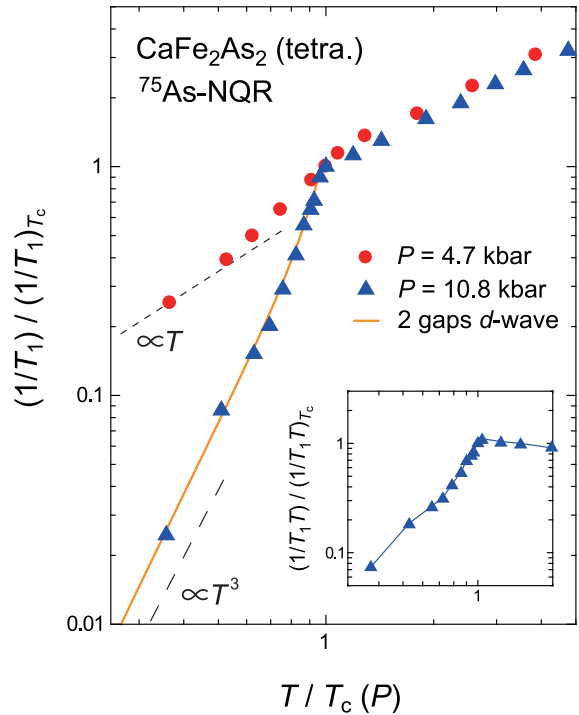}

\caption{\label{fig:t1t} (Color online) Plot of $T_1(T)$$^{-1}$/$T_1(T_c)$$^{-1}$ versus $T/T_c$$(P)$ for $P$ = 4.7 and 10.8 kbar. The solid curve is a two-gap fit assuming a $d$-wave symmetry with parameters, $\Delta_1(0) = 3.9 k_{B}T_{c}$, $\Delta_2(0) = 1.7 k_{B}T_{c}$,  and  $\alpha = 0.65$ (see text). The dotted and dashed lines indicate the relations of $1/T_1$ $\propto$ $T$ and $1/T_1$ $\propto$ $T^3$, respectively. Inset shows plot of ($T_1T(T)$)$^{-1}$/($T_1T(T_c)$)$^{-1}$ versus $T/T_c$$(P)$ for 10.8 kbar. Solid line is eye-guide.}
\end{center}
\end{figure}

To focus on the superconducting gap structure for CaFe$_2$As$_2$, the plots of $T_1(T)$$^{-1}$/$T_1(T_c)$$^{-1}$ versus $T/T_c$$(P)$ are shown in Fig. 5. Here, the relaxation rate below $T_{c}$ ($1/T_{1s}$) can be expressed as, $\frac{T_{1N}}{T_{1s}}= \frac{2}{k_BT} \int \int N_s(E)N_s(E')f(E) \left[1-f(E') \right] \delta(E-E')dEdE'$. Where $Ns=\frac{E}{\sqrt{E^2- \Delta^2}}$ is the DOS in the superconducting state, and $f(E)$ is the Fermi distribution function. The coherence peak just below $T_c(P)$ is absent at both pressures. At $P$ = 4.7 kbar, $1/T_1$ decreases moderately and is saturated approaching $T$ = 0.  This  means that there is a residual density of states in the superconducting gap.  Since the present experiments are performed in zero magnetic field, it is clear evidence for the occurrence of gapless superconductivity.

On the other hand, at $P$ = 10.8 kbar, $1/T_1$ continues to decrease steeply below $T_c$, as observed in other iron pnictide superconductors\cite{MatanoPr,KawasakiLa,MatanoBa122,Nakai,Mukuda,Grafe,KotegawaFeSe,Fukazawa,Kobayashi}. Notably, as clearly seen in the inset to Fig.5, $1/T_1T$ below $T_c$ has a hump structure around $T$ $\sim$ 0.5 $T_c$, which is a signature for  multiple gap superconductivity, as observed in other pnictide superconductors\cite{MatanoPr,KawasakiLa,MatanoBa122}.  By assuming two gaps of $d$-wave symmetry $\Delta(\phi) = \Delta_0 \cos(2\phi)$ with a mean-field temperature dependence with $\Delta(\phi)=\alpha \Delta_1+ (1-\alpha) \Delta_2$ and $\alpha = \frac{N_{s,1}}{N_{s,1}+N_{s,2}}$, we find that the $\Delta_1(0) = 3.9 k_{B}T_{c}$, $\Delta_2(0) = 1.7 k_{B}T_{c}$  and  $\alpha = 0.65$ can fit the data reasonably well, as shown by the solid curve in Fig.5. These values of superconducting gaps and $\alpha$ are comparable to other iron pnictide superconductors\cite{MatanoPr,KawasakiLa,MatanoBa122}.  
    
How can we understand this difference of the superconducting gap structure between $P$ = 4.7 and 10.8 kbar? 
 One possible scenario is the mechanism predicted in heavy fermion superconductivity around the antiferromagnetic QCP at which the gapless superconductivity has been observed\cite{YuKawasaki,Kawasaki115,Yamaguchi}. When the system locates at the vicinity of the antiferromagnetic QCP, odd-frequency $p$-wave spin singlet superconductivity (pSS) prevails over the $d$-wave singlet superconductivity (dSS)\cite{Fuseya}. Notably, for the pSS state, it is suggested that there is no gap in the quasiparticle spectrum anywhere on the Fermi surface due to its odd frequency, thus, gapless superconductivity is realized\cite{Fuseya}. In the present case, the values of $\theta$, which is the measure of closeness to the QCP are very small and  comparable to the value of heavy fermion compounds around QCP.\cite{ZhengIr,KawasakiCeRhIrIn5} In this model, gapless pSS and dSS compete near a QCP.\cite{Fuseya} In addition, it would be difficult to realize gapless pSS when it competes against full-gap superconductivity, such as $\pm$$s$-wave pairing\cite{Fuseya2}. Thus, $d$-wave pairing is favored as the competing order against gapless pSS near a QCP\cite{Fuseya2}. In fact, it has been predicted that $d$-wave superconductivity can also be the candidate for iron-pnictide superconductivity, although $\pm$$s$-wave pairing has been suggested in other iron pnictide superconductors\cite{Kuroki}. The present results suggest that the pressure-induced superconductivity near a QCP in CaFe$_2$As$_2$ is a  good candidate to investigate a variety of superconductivities in iron pnictides.

\section{Concluding remarks}
In conclusion, we report zero-field NMR/NQR experiments on the iron-pnictide pressure-induced superconductor CaFe$_2$As$_2$. Systematic measurements have revealed the evolution of the ground states under pressure and the electron correlations play a vital role in the  formation of Cooper pairs in this compound. It is suggested that the electron correlation is induced by an  interband correlation originating from its multiple bands structure. We found that gapless superconductivity is realized in the close vicinity of the  antiferromagnetic quantum critical point. We believe that it is due to its closeness to a quantum critical point. The present results suggest a close relationship between antiferromagnetism and superconductivity in iron pnictides.\\

\section*{Acknowledgement}
S. K. thanks Yuki Fuseya for useful discussion and comments. This work was supported by a Grant-in-Aid for Scientific Research on Innovative Areas "Heavy Electrons" (No. 21102514)  of The Ministry of Education, Culture, Sports, Science, and Technology, Japan.


\end{document}